# Electrode/electrolyte Interface in the Li-O$_2$ Battery: Insight from Molecular Dynamics Study.

Artem Sergeev[1], Alexander Chertovich[1*], Daniil Itkis[2], Anik Sen[3], Axel Gross[3] and Alexei Khokhlov[1]

[1] Physics Department, Lomonosov Moscow State University, Moscow, 119991 Russia

[2] Chemistry Department, Lomonosov Moscow State University, Moscow, 119991 Russia

[3] Institute of Theoretical Chemistry, Ulm University, Ulm, 89081 Germany

**Abstract**

In this paper for the first time we report the results of molecular dynamics simulation of electrode/electrolyte interface of Li-O$_2$ cathode under potential close to experimental values in 1M dimethyl sulfoxide (DMSO) solution of LiPF$_6$ salt. Electric potential profiles, solvent structuring near the electrode surface and salt ions distributions are presented and discussed here as well as potentials of mean force (PMF) of oxygen and its reduction products. The latter would be of a great use for the future theoretical studies of reaction kinetics as PMF being essentially the work term is a required input for the reaction rate constant estimations. Electrode/electrolyte interface under the realistic potential effectively push oxygen anions out of the reaction layer that makes second reduction of superoxide anion hardly probable. Thus the main cause of the passivation should be the lithium superoxide presence near the electrode surface. The way to suppress the passivation is the shifting of equilibrium $O_2^- + Li_+ \rightleftharpoons LiO_2$ to the side of separately solvated ions, for example by using solvents resulting in lower free energy of the ions. This conclusion is in agreement with the hypothesis stating that high donor number solvents lead to dominantly solution Li$_2$O$_2$ growth and significantly higher cell discharge capacities.

**Introduction**

Combination of alkali metals with the one of the strongest yet lightest oxidizer promises enormous benefits for the energy storage and conversion. For several decades it has been alluring researches into studying metal-oxygen redox systems. Li-O$_2$ cells being an implementation of this idea might become the next generation of battery technology. However, there is a list of practical issues including low operation current and voltage efficiency, poor cycle life and cell capacity which lies far below the expectations. The fundamental understanding of oxygen reduction mechanisms in Li-salt containing media is necessary to overcome these challenges.

High overvoltage on discharge and low cell capacity are usually attributed to passivation of electrode surface [1–4] that severely inhibits and eventually stops the electrochemical reaction. Passivation can be caused by degradation of solvent and electrode due to attacks of O$_2^-$ intermediate [1,5] that hopefully may be avoided by the proper choice of the cell materials. Unfortunately the cell discharge product Li$_2$O$_2$ itself is known to be an insulator with only few crystal surfaces reported to exhibit conducting properties according to DFT calculations [6,7]. Thus passivation problem stays essential for Li-O$_2$ systems.

---

* Corresponding author: chertov@polly.phys.msu.ru

Two mechanisms of $Li_2O_2$ formation is proposed in the literature, see Figure 1. The first one suggests oxygen reduction to $O_2^-$, association with $Li^+$ ion and subsequent disproportionation of lithium superoxide yielding lithium peroxide.[8] This pathway should mostly lead to generation of $Li_2O_2$ in solution and formation of large micron-sized particles.[9,10] On the other hand, lithium peroxide can be result of second reduction of superoxide species that would lead to surface product formation and passivation of the electrode.[4,11] The later mechanism is off course undesirable as it severely limits cell capacity. There are also attempts to develop a universal description of $Li_2O_2$ formation that includes both pathways.[12,13]

Superoxide anion is known to be a direct product of electrochemical oxygen reduction.[14,15] Thus $O_2^-$ should be present in Li-$O_2$ cell at least as a short-lifetime intermediate. Bruce and coworkers[13] suggest that separately solvated $O_2^-$ and $Li^+$ can exist in solution in thermodynamic equilibrium with molecular $LiO_2$. Therefore, one should take into account the possibility of second reduction of $O_2^-$ to $O_2^{2-}$.

An ultimate understanding of the system would mean the creation of a model that include rate constants of all reaction steps and is able to provide spatial distributions of reaction rates and concentrations of species near the electrode surface. Although for now this goal seems unreachable and even excessive, any information about the structure of the electrode/solution interface would be indeed useful for the further research and development of LI-$O_2$ batteries.

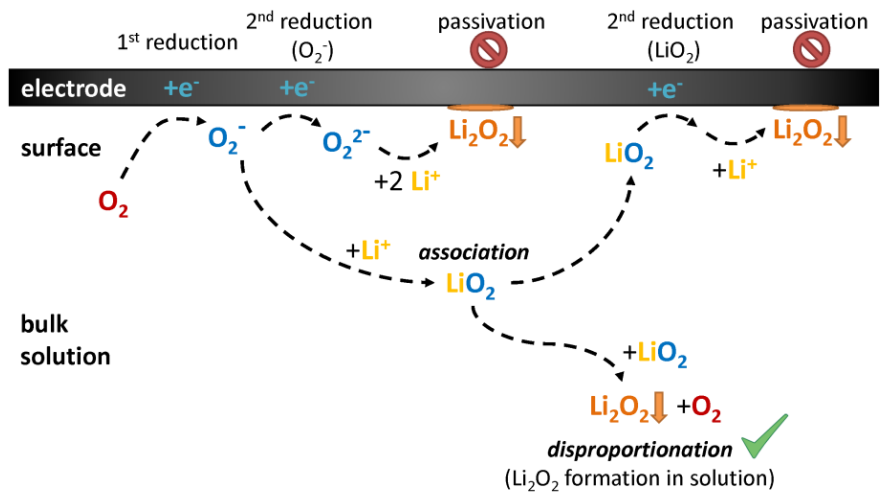

Figure 1: A schematic representation of the possible reaction steps.

Atomistic computer simulations proved to be an productive tool for investigation of an electrode/solution interfaces[16,17]. Simulation of an electrochemical interface is a challenging task for DFT methods. Nevertheless some promising approaches to this problem have been proposed in literatute.[18] Unfortunately, due to high computational costs only small systems with solvent taken into account implicitly are feasible. Structure of large (in comparison to water) organic solvent molecules would have a great influence on the solution organization near the electrode surface, while concentrations of the species changes across a layer up to 10 angstroms thick or even more[19,20]. Although classical molecular dynamics lacks the description of quantum phenomena it seems to be a reasonable compromise between accuracy and scale (both time and spatial) for the electrode/electrolyte interface simulation. The trickiest part of electrochemical interface simulation may be an adequate reproduction of reasonable electrode potential. One possible solution is to use potential of zero charge (p.z.c.) as a reference value between simulations and experiments.

**Computational details**

All molecular dynamics simulations are performed using the Large-Scale Atomic/Molecular Massively Parallel Simulator (LAMMPS) package[23]. VMD software[24] was used for visualization and postprocessing of simulation data including calculation of electric potential profiles by means of PME electrostatics extension[25]. The simulation cell was designed as a parallel-plate capacitor (FIG 2) filled with electrolyte solution. Plates were perpendicular to z axis and their x/y dimensions are fixed to 29.30 X 30.45 Å. The fixed distance between plates is 80 Å. It was chosen to reproduce density obtained from bulk solution simulation (1.176±0.006 g/cm$^3$) in the 20 Å thick middle region of the cell.

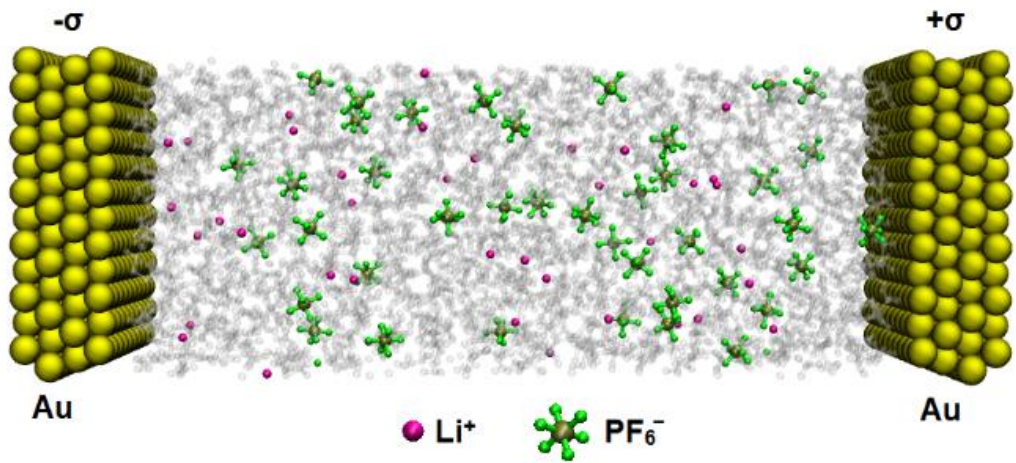

Figure 2: A simulation cell snapshot. Solvent molecules are in ghost-grey. Solution region is 80 Å wide. σ = 9.7 μC/cm$^2$.

Periodic boundary conditions was applied, but special technique provided by LAMMPS package was used to eliminate unwanted interactions between the replicas in z direction [26,27]. That included insertion of 220 Å vacuum region and subtracting of electrostatic interaction between the replicas. Long range correction (particle-particle-particle-mesh (pppm)[28]) was used to calculate Columbic interactions beyond 12 Å cutoff. Van der Waals interactions was also calculated with 12 Å cutoff. All productive runs were carried out at 298 K in NVT ensemble using Nose-Hoover thermostat.

Solution consisted of 512 molecules of DMSO, 40 Li$^+$ and 40 PF$_6^-$ ions that correspond to about 1M salt concertation. A flexible all-atom model of DMSO developed by Strader and Feller[29] was adopted. Force field parameter of Li$^+$ and PF$_6^-$ ions designed for organic solution was taken from recent work by Kumar and Seminario[30]. Lorentz−Berthelot mixing rules were applied. To describe the electrode we chose GolP model[31,32] of Au(111) surface that efficiently takes into account polarization (image charge) effects. Gold plates were composed of four atomic layers. In case of non-zero electrode charge small charge was added to every surface gold atom of the plate and balanced by opposite the charge at the other plate.

Potential of mean force calculations was carried out for a list of reagents: $O_2$, $O_2^-$, $LiO_2$, $Li^+$. Averaged force acting on the molecule of interest was calculated at a set of distances from the electrode surface (the plane containing the centers of gold atoms composing the first layer). At every distance the molecule have been bound by a harmonic potential (k = 1500 kcal/mol/Å$^2$) to the certain plane parallel to the electrode surface. On average, net force exerted by the surroundings should be balanced by the force of the harmonic bond. So the force exerted by the bond was used to calculate the mean force acting on the molecule. At every distance system was equilibrated for 0.5 ns and then the force was averaged during 1 ns simulation. After that the molecule was dragged to another distance. The planes were located in the range of distances from 3.2 Å to 20 Å from the surface. The spatial step was 0.5 Å for distances greater than 8 Å and 0.2 Å for smaller distances. At least two runs were performed: forward (moving the molecule toward the surface) and backward. The averaged force data was than interpolated and integrated to obtain the PMF.

The O-O bond length of the $O_2$ and $O_2^-$ molecules was set to 1.21 Å in accordance to experimental data[33]. The stretching force constant set to 1694 kcal/mol/Å$^2$ have been derived from the experimental vibrational frequencies [34]. Lennard-Jones parameters for oxygen atom in $O_2$, $O_2^-$, $LiO_2$ have been taken same as the CHARMM amide oxygen. Lithium LJ parameters for $LiO_2$ molecule was the same as for $Li^+$ ion.

Geometry and partial charges of $LiO_2$ molecule have been obtained with the help of DFT calculations using Gaussian 09 software[35]. The geometry optimization was carried out using B3LYP functional and 6-31++G** basis set followed by single point and frequency calculations with aug-cc-pVTZ basis set. Partial atomic charges were fit to the electrostatic potential. Li-O bond constants were derived from the frequencies analysis. Two sets of parameters were prepared (see Table 1) based on vacuum calculations ($LiO_{2\_vac}$) and implicit solvent calculation using IEFPCM scheme ($LiO_{2\_DMSO}$).

**Table 1: Set of LiO$_2$ molecule parameters derived from DFT calculations and later used in MD simulations.**

|  | $LiO_{2\_vac}$ | $LiO_{2\_DMSO}$ |
|---|---|---|
| O charge, $e_0$ | +0.26 | +0.90 |
| Li charge, $e_0$ | -0.48 | -0.45 |

**Results and discussions**

Electrode potential has a great influence on the structure of the electrode/solution interface[19,36]. According to experimental data[37] p.z.c. of gold electrode in DMSO is +0.39 V vs. SHE (+3.43 V vs. Li$^+$/Li) while standard potential of $O_2/O_2^-$ is reportedly -0.78 V vs. SCE[38] (+2.50 V vs. Li$^+$/Li) and common discharge potential range of Li-$O_2$ cells is 2.4-2.8 V vs. Li$^+$/Li. As the potential of interest ("operation potential" here) is about 1 V below the p.z.c. that would be unjustified to neglect the potential aspect in the simulations.

Therefore, we ran a series of simulations varying charge of the plates and calculated electric potential profiles across the cell (Fig 3). Negative charge up to -0.05 $e_0$ was added to every surface gold atom of the left (lower z) plate. Electric potential converges to its bulk value within 3 molecular layers (≈15 Å). Surface charge -σ = -0.045 e0/Au = -9.7 μC/cm$^2$ yields electrode potential about 1 V below the potential of bulk solution. That corresponds to the difference of p.z.c. and operation potential in the experiments. This surface charge value later referred as "proper" was used for PMF calculations.

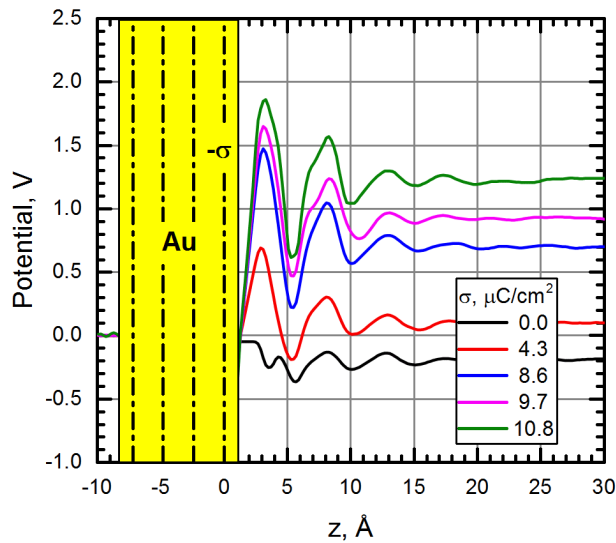

Figure 3: Electric potential profiles near the left electrode at different negative surface charge values. Atomic planes of the gold plate are indicated by dash-dot lines.

To evaluate the influence of the electrode charge we have compared solvent structuring (Fig 4) and salt ions distributions (Fig 5) in cases of zero and proper negative surface charge. In fig 4 solvent concentration (based center of mass coordinates) normalized by bulk value is depicted by dashed black line. Oscillation of solvent concentration near the surface is a common behavior and there is no significant difference between charged and uncharged cases. Orientational order parameters $P_1 = \langle cos(\theta) \rangle$ and $P_2 = \langle 3/2 cos^2(\theta) - 1/2 \rangle$, where $\theta$ is the angle between z direction and normal to the plane containing S and C atoms of DMSO molecule, are also depicted in fig. 4. In case of proper negative surface charge (fig. 4B) one can observe prominent orientation of solvent molecules ($\langle \theta \rangle \to 0$) in such a way that S atoms and $CH_3$ groups lie on the electrode surface while O atoms point out toward the solution (fig. 4C). The orientation is much weaker near the non-charged surface (fig. 4A). Thus, solvent ordering in the first molecular layer is at least partially originate from coulomb interaction of the charged surface with positively charged S atoms and $CH_3$ groups and negatively charged O atoms. The effect is important for electrochemical reactions as preoriented solvent molecules should lose the ability to effectively solvate ions.

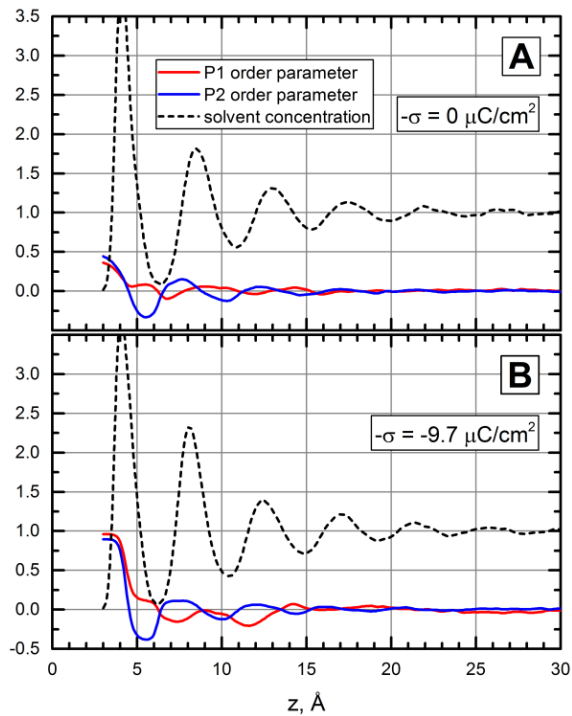

Figure 4: Order parameters (red and blue solid) and normalized solvent concentration (black dashed) profiles near the neutral (A) and negatively charged (B) surfaces. C - explanatory sketches.

The ions distributions are compared in fig 5. There are only small oscillations of ions concentrations near the neutral surface. Note that there is a substantial concentration of salt ions closer than 5 Å to the surface, that is right inside the thirst molecular layer. (About 3 Å wide layer is abandoned due to Pauli repulsion (modeled with Lenard-Jones potential) from Au atoms situated accurately at z = 0.) In case of negative surface charge a high peak of $Li^+$ concentration can be observed near the electrode. It is located at z = 6 Å, just between first and second molecular layers of solvent (see DMSO concentration peaks in Fig 4), rather than next to the negatively charged surface, as one could expect relying on Coulomb attraction. $PF_6^-$ anions are almost completely pushed out beyond second molecular layer of solvent.

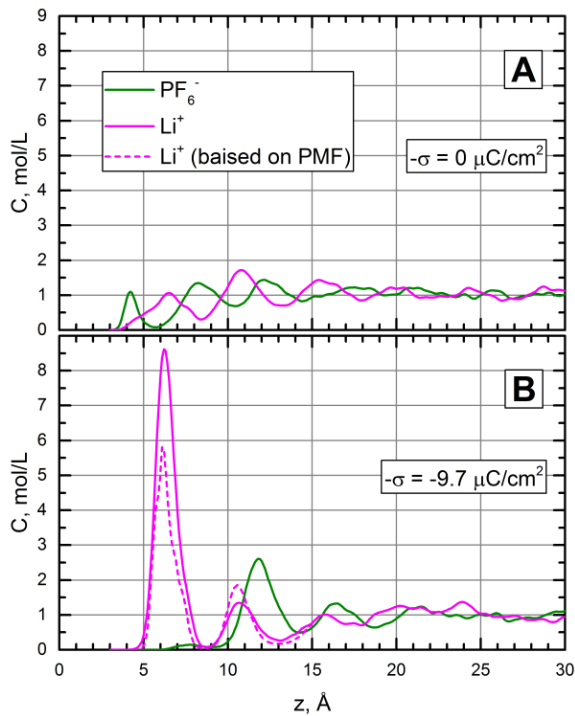

Figure 5. Salt ions concentration profiles near the neutral (A) and negatively charged (B) surfaces.

We would like to highlight as the first conclusion of current work that the negative surface charge at the electrode operation potential cause significant structural changes (in comparison to zero surface charge) of the electrode/electrolyte interface of the Li/O$_2$ cathode with DMSO based solution. Besides expected attraction of salt cations and repulsion of anions, a complete exclusion of all ions (irrespectively of polarity) from the first molecular layer of DMSO is predicted. It is accompanied by prominent orientation of DMSO molecules of the first layer. Thus taking into account electrode operation potential in simulations of Li/O$_2$ cathode is absolutely necessary to obtain relevant results.

Before discussing the PMF results we remind that according to our DFT calculations (see Table 1) partial charges of lithium superoxide molecule strongly depend on dielectric permittivity of the medium. In vacuum calculations showed that associating O$_2^-$ and Li$^+$ ions share almost half of its charge with the partner significantly reducing the overall dipole moment. When surrounding solvent was taken into account implicitly by means of polarizable continuum model (PCM), ions retained most of its initial charge. PCM is a good approximation for solvation in bulk liquid but dielectric properties of solvent changes[39] at the solid/liquid interface. It is reasonable to assume that the lack of orientational freedom of the DMSO molecules in the first layer weakens the ability to screen ion charge, i.e. lowers local dielectric permittivity. Therefore, we performed PMF calculations using two sets of LiO$_2$ parameters based on vacuum (LiO$_{2\_vac}$) and implicit solvent (LiO$_{2\_DMSO}$) DFT calculations expecting the truth to be between these two limiting cases.

The calculated PMF profiles are presented in fig. 6. Neutral O$_2$ molecule, can reach the very surface by overcoming the free energy barrier about 2.5 k$_b$T. PMF of O$_2^-$ and Li$^+$ exhibit a steep rise starting from about in between first and second molecular layers (z = 6 Å). That confirms that intercalation of ions in the first DMSO layer is improbable. The same is true for the LiO$_{2\_DMSO}$ being a bound pair of ions *per se*.

LiO$_{2\_DMSO}$ experiences weaker repulsion from the surface in comparison to LiO$_{2\_DMSO}$ version due to reduced atomic charges. PMF of O$_2^-$ near the neutral surface (dashed blue line) does not rise high. Thus, ions exclusion from the first molecular layer caused by negative charge of the electrode surface.

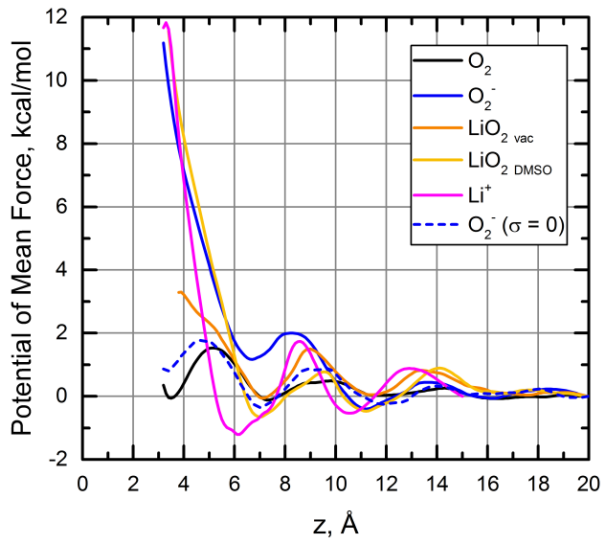

Figure 6. PMF of the reagents near the negatively charge surface (solid lines) and PMF of superoxide anion near the neutral surface (dashed line).

As according to the results free energy of both superoxide anion and lithium superoxide in the first molecular layer is extremely high we conclude that adsorption and consequent disproportionation of these species directly on the surface is doubtful.

Boltzman distribution was employed to estimate concentration profiles of reagents near the electrode surface ($c_X(z)/c_{X,bulk} = exp(PMF_X(z)/k_bT)$), as direct calculations are impossible due to very low concentration (~ 1 mM, less than 1 molecule per simulation cell). First we validated the approach by comparing PMF based Li$^+$ concentration profile with the directly calculated (fig. 5B, dashed line). A good quantitative agreement was achieved.

Fig. 7 represents reagent concentration profiles normalized by bulk value. The most important feature is that there are substantial peaks of lithium superoxide (for both LiO$_{2\_vac}$ and LiO$_{2\_DMSO}$ parameter sets) located around z = 7 Å, while O$_2^-$ concentration vanishes at distances smaller than 10 Å from the surface. Electron transfer probability decays exponentially with distance $\kappa_{et}(z) = \kappa_0 exp(-z/\lambda_{et})$ and the characteristic length $\lambda_{et}$ is about 1 Å[21,40]. Thus presence of lithium superoxide concentration peak at a 3 Å closer distance favors LiO$_2$ reduction rather than second reduction of superoxide anion ($O_2^- + e^- \rightleftharpoons O_2^{2-}$) from the kinetic point of view.

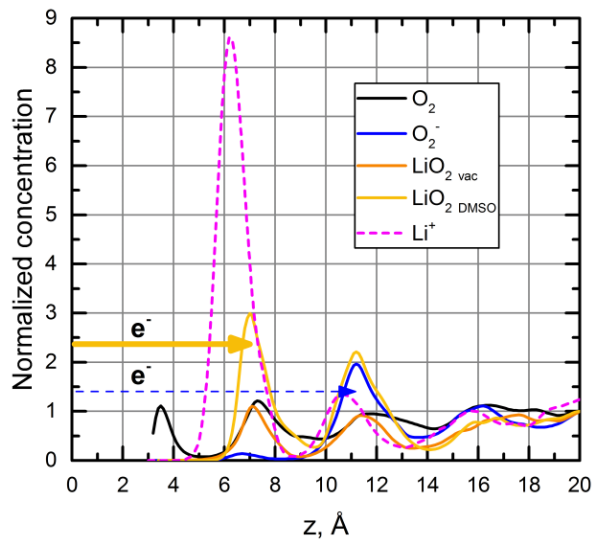

Figure 7. Concentration profiles of the reagents calculated using PMF (solid lines). Li$^+$ concentration profile (dashed line) obtained directly from the simulation. Electron transfer probability decays with distance from the electrode.

Lithium salt is known to reduce $O_2/O_2^-$ reduction potential. That provides us ground to assume that reaction free energy of $LiO_2$ reduction is lower than that of $O_2^-/O_2^{2-}$ reduction. Thus both kinetically and thermodynamically second reduction of superoxide anion is less favorable than lithium superoxide reduction. The conclusion is that second electron reduction only becomes possible after association of superoxide anion with lithium cation.

This last conclusion should be discussed in more details. Electrochemically generated superoxide anions diffuse away from the electrode and convert into lithium superoxide on the way. The higher the association rate the higher the concentration of lithium superoxide near the electrode surface that promotes second electron reduction leading to passivation. We suggest, that inhibition of association reaction should decrease $LiO_2$ concentration and, consequently, diminish the passivation rate.

Thermodynamic way to inhibit a reaction is to make its free energy difference $\Delta G = G_{product} - G_{reagents}$ more positive, for example, by decreasing free energy of the reagents. Therefore, solvents demonstrating more negative solvation energies of associating ions (i.e. higher donor and acceptor numbers) should make association less favorable and promote solution phase lithium peroxide formation. This inference is in agreement with the experimentally backed hypothesis[11,13] stating that high donor number solvents lead to dominantly solution phase $Li_2O_2$ growth and significantly higher cell discharge capacities.

Another way to suppress passivation can be suggested basing on the obtained results. Note the high Li$^+$ concentration peak (Fig 5B) near the electrode surface, just where $O_2^-$ is generated. It arises in response to negative surface charge and screens it. The abundance of lithium ions apparently leads undesirable $LiO_2$ formation and should be eliminated. We propose that the lithium ions could be replaced by chemically inactive cations with high adsorption energy to the electrode surface. Such cations introduced into solution would scree negative charge of the surface instead of lithium ions. Removing extra lithium ions from the interface should substantially increase oxygen superoxide lifetime and reduce lithium superoxide concentration near the surface hence suppressing the passivation process.

Finally, we have to discuss limitations of the approach used. The results and particularly strong ordering of first solvent layer are obtained for the ideally flat electrode surface. It should still stand for surfaces with curvature radius much greater than the molecular layer thickness ($\approx 5$ Å). However, additional study should be conducted on the influence of surface defects and angstrom-scale roughness. Some relevant attempts have been made recently[41], unfortunately electrode potential (and/or charge) was not taken into account, which is of high importance for modeling Li-$O_2$ system according to our findings. Also we do understand that adsorption of DMSO molecules on the electrode surface can be influenced by quantum effects that are not captured by molecular dynamics simulation. Therefore, our future plans include periodic DFT calculations of single molecular layer of DMSO on the electrode surface under the proper potential.

**Conclusions**

All-atom molecular dynamics simulation of the interface between gold electrode and 1M DMSO solution of $LiPF_6$ salt were carried out under realistic conditions. Polarizable model of Au (111) surface was employed to allow for image charge effects. Potential of zero charge was used as a reference to find the value of the surface charge that corresponds to cathode operation potential about +2.4-2.6 V vs $Li^+$/Li. Solvent structure at the interface, concentration profiles of salt ions and reagent species and values of mean force were calculated at correct cathode potentials.

Negative surface charge was found to cause significant changes of the interface structure. Particularly, different ions redistribution and specific orientation of solvent molecules in the first layer was observed near the electrode surface. Such changes would have a great influence on the electrochemical and chemical reactions in the system. Therefore, we claim it necessary for the atomistic simulations of Li-$O_2$ systems to take the electrode operation potential into account.

Directly calculated concentration and PMF profiles show complete expulsion of all ions (regardless of polarity) from the first molecular layer. Highly polar $LiO_2$ molecule is also a subject to this expulsion. Thus we find that adsorption and further disproportionation of superoxide anions and lithium superoxides directly on the surface is unlikely.

The value of the negative surface charge at the electrode operation potential turned out to be enough to push $O_2^-$ beyond 10 Å limit from the electrode surface, which suppress the electron transfer from the electrode. On the other hand, lithium superoxide concentration peak predicted to be at 7 Å distance that enables effective electron transfer. These findings support that second electron reduction take place only for lithium superoxide, i.e. only after the association of $O_2^-$ and $Li^+$ have occurred.